\documentclass[12pt]{iopart}
\usepackage{epsfig}
\begin{document}
\jl{1}

\title[Symmetry group analysis of relativistic fluids]
{Symmetry group analysis of relativistic heat conducting fluids}

\author{C Alexa}

\address{Particle Physics Dept., IFIN-HH, Bucharest 76900, Romania}

\begin{abstract}
The Lie symmetry group for 1+1 dimensional relativistic heat-conducting 
fluid is calculated for two different theories, Eckart and Israel-Stewart  
and a comparison between the group-invariant solutions has been made.
Both fluids were founded to be physical acceptable in the sense that 
during the evolution of the fluid there are three velocity solutions 
that are decreasing exponentially for particular choices of the initial 
conditions.
\end{abstract}

\submitted
\maketitle

\section{Introduction}

Relativistic fluid dynamics provide a simple but intuitive description of 
many physical systems. The hot and compressed nuclear matter behaves like a 
compressible fluid and fluid dynamical effects are observed in high energy 
heavy ion reactions \cite{csernai}. 
Supernova or neutron star may be another situation 
where the relativistic heat-conducting fluid gives a satisfactory 
description.

The standard theory of the relativistic dissipative fluid proposed by 
Eckart \cite{eckart} and a similar one by Landau and Lifshitz \cite{landau} 
was analysed by Hiscock and Lindblom \cite{h1,h2,h3,h4} and it
was found to be unstable, acausal and ill posed in the linear regime near
equilibrium.
A more complicated extended hydrodynamic theory has been proposed by Israel 
and Stewart \cite{i1,i2,i3} which is free of the most part of the troubles
of the standard theory. 

The Hiscock and Lindblom studies were dedicated to 
causality, stability, hyperbolicity and to the connection between them.
In \cite{h4} the solution discussion was restricted to spatially homogeneous
fluid. This paper will improve the dimensional restriction 
considering the evolution equations for the plane symmetric motion
(1+1 dimensional) of the relativistic heat-conducting fluid and 
propose a complementary analysis dedicated to group-invariant 
solutions study.

Symmetry analysis is a systematic and accurate way to obtain solutions of
differential equations. The Lie symmetry group method is well known but the 
major obstacle in the application of this method to partially differential 
equations is the large number of tedious calculations usually involved. 

We begin our analysis by the calculation of the Lie symmetry group for
both Eckart and Israel-Stewart fluids using, for the equation of state, the 
high temperature limit of the ideal gas, where $p = \rho / 3 = n k T$.
This constraint allows us to eliminate from the fluid equations two dependent 
variables, in particular, we substitute the temperature $T$ and the pressure
$p$ in favour of the energy density $\rho$ and the number density $n$.
Next, we make a systematic classification of the solutions 
constructing the optimal system of transformations. 
The investigation of the solutions is completed by solving the reduced 
system of equations; we present both analytical and numerical solutions.

\Sref{sfl} will display the system of equations of the relativistic 
dissipative fluid in its general form eqs. \eref{gf} and in the 
particular case of heat-conducting theory eqs. \eref{eqs}. 
The application of the Lie symmetry group method to differential equations 
and the corresponding Lie algebra of the fluid equations \eref{eqs}  
are presented  in \Sref{lie}. \Sref{inv} is devoted to integrability 
conditions and invariants, \Sref{clas} to the classification of the solutions 
and \Sref{sol} to group invariant solutions analysis.
Possible applications of the group-invariant solutions to high energy
collisions, miscellaneous comments and final conclusions are discussed in 
\Sref{sum}.  

\section{Dissipative relativistic fluid mechanics}\label{sfl}

In the particle frame, used by Eckart, the energy tensor $T^{\alpha\beta}$ 
and the particle flux vector $N^{\alpha}$ take the form \cite{i1}:
\begin{equation}\label{gf}
\eqalign{
T^{\alpha\beta} & =  \rho u^{\alpha} u^{\beta} 
+ (p+\sigma)\left( g^{\alpha\beta} + u^{\alpha} u^{\beta} \right)
+ q^{\alpha} u^{\beta} + q^{\beta} u^{\alpha} + \sigma^{\alpha\beta} \\
N^{\alpha} &  = n u^{\alpha}
}
\end{equation}
where $u^{\alpha}$ is a unit time-like vector field which may be thought of as
the four-velocity of the fluid, $g^{\alpha\beta}$ is the Minkowski metric 
diag(-1,1,1,1), $\rho$ the energy density of the fluid, 
$p$ the thermodynamical pressure, $n$ the number density,
$q^{\alpha}$ is the heat flow, $\sigma$ the bulk stress and 
$\sigma^{\alpha\beta}$ the shear stress.

The heat conducting case is obtained by setting $\sigma = 0$ 
and $\sigma^{\alpha\beta} = 0$. In this particular condition 
the heat flow obeys the following phenomenological law \cite{i1}:
\begin{equation}\label{qal}
q^{\alpha} = - \kappa T \left( g^{\alpha\beta} + u^{\alpha} u^{\beta} \right)
\left( T^{-1}\partial_{\beta}T + u^{\mu}\partial_{\mu}u_{\beta} 
+\bar\beta_1 u^{\mu}\partial_{\mu} q_{\beta} \right)
\end{equation}
where beside the usual thermal conductivity $\kappa$ and temperature $T$ 
we have a new phenomenological coefficient $\bar\beta_1$ \cite{i1}. Setting 
$\bar\beta_1 = 0$ we obtain the standard Eckart heat conducting fluid theory. 

The conservation laws:
\begin{equation}\label{cons}
\partial_{\alpha} T^{\alpha\beta} = 0,\>\partial_{\alpha} N^{\alpha} = 0 
\end{equation}
and the equation \eref{qal} form the complete system of equations of the 
general theory of relativistic dissipative fluids.

If we write the four-velocity in the form 
$u^{\alpha}=\left(\cosh\Psi , \sinh\Psi , 0 , 0\right)$, 
the heat flow vector, which is orthogonal to $u^{\alpha}$, will be  
$q^{\alpha} = q\left(\sinh\Psi, \cosh\Psi, 0, 0\right)$ and 
the system of equations of the fluid becomes:
\begin{equation}\label{eqs}
\begin{array}{ll}
\fl \sinh\Psi \partial_x n - \cosh\Psi \partial_t n 
+ n \cosh\Psi \partial_x\Psi - n \sinh\Psi \partial_t\Psi = 0&  \\
\fl \cosh\Psi \partial_t\rho - \sinh\Psi \partial_x\rho 
+ \sinh\Psi \partial_t q - \cosh\Psi \partial_x q \\
\fl + \left[\left(p+\rho\right)\sinh\Psi + 2 q \cosh\Psi\right]\partial_t\Psi
- \left[\left(p+\rho\right)\cosh\Psi + 2 q \sinh\Psi\right]\partial_x\Psi
= 0&  \\
\fl \cosh\Psi \partial_x p - \sinh\Psi \partial_t p 
+ \sinh\Psi \partial_x q - \cosh\Psi \partial_t q  \\
\fl - \left[\left(p+\rho\right)\cosh\Psi + 2 q \sinh\Psi\right]\partial_t\Psi
+ \left[\left(p+\rho\right)\sinh\Psi + 2 q \cosh\Psi\right]\partial_x\Psi
= 0& \\
\fl T^{-1}\left(\cosh\Psi \partial_x T - \sinh\Psi \partial_t T \right)
+ \bar\beta_1 \left(\sinh\Psi \partial_x q - \cosh\Psi \partial_t q \right) \\
\fl + \sinh\Psi \partial_x\Psi - \cosh\Psi \partial_t\Psi 
+ \frac{q}{\kappa T} = 0& 
\end{array}
\end{equation}
We will supplement the eqs. \eref{eqs} by an equation of state (EOS) for
the fluid and a thermodynamic expression for $\bar\beta_1$.
Choosing the high temperature limit of an ideal gas, the equation of state is
\begin{equation}\label{eos}
p = \rho / 3 = n k T
\end{equation}
Finally, we take the thermal conductivity to be constant and the second-order 
thermodynamic coefficient $\bar\beta_1$ to be given by 
$\bar\beta_1 = 5 \lambda / 4 p$; the parameter $\lambda$ will be $1$ for
the Israel-Stewart theory and $0$ for the Eckart theory. 

The equation \eref{eos} allows us to eliminate $T$ and $p$ in favour of 
$\rho$ and $n$, obtaining finally a closed set of evolution equations
for four variables $\left(\Psi, n, \rho, q\right)$.

\section{Symmetry group of transformations and its Lie algebra}\label{lie}
The symmetry group of a system of differential equations is the largest
local group of transformations acting on the independent and dependent
variables of the system with the property that it transforms solutions of the
system to other solutions. 
Let ${\cal S}$ be a system of differential equations. A symmetry-group of
the system ${\cal S}$ is a local group of transformations ${\cal G}$ acting
on an open subset ${\cal M}$ of the space of independent and dependent
variables of the system with the property that whenever u=f(x) is a
solution of ${\cal S}$, and whenever $g\cdot f$ is defined for $g\in {\cal G}
$, then $u=g\cdot f(x)$ is also a solution of the system.

The symmetry group infinitesimal generator is defined by : 
\begin{equation}
\vec {{\cal V}}=\tau\partial_t+\xi\partial_x+\Phi\partial_{\Psi}
+\Sigma\partial_n+\Gamma \partial_{\rho}+\Omega \partial_q 
\end{equation}
and the first order prolongation of $\vec {{\cal V}}$ is: 
\begin{equation}
\begin{array}{ll}
pr^{(1)}\vec {{\cal V}}= & \xi \partial _x+\tau \partial _t
+\Phi \partial_{\Psi}+\Sigma \partial_n+\Gamma \partial_{\rho}
+\Omega \partial_q \\  
& +\Phi^x\partial_{\Psi_x}+\Phi^t\partial_{\Psi_t}
+\Sigma^x\partial _{n_x}+\Sigma^t\partial_{n_t} \\  
& +\Gamma^x\partial_{\rho_x}+\Gamma^t\partial_{\rho_t}
+\Omega^x\partial_{q_x}+\Omega^t\partial_{q_t} 
\end{array}
\end{equation}
where, for example, 
\begin{equation}
\Phi^x=D_x(\Phi -\xi\Psi_x-\tau\Psi_t)+\xi\Psi_{xx}+\tau\Psi_{xt} 
\end{equation}
and $D_x\Phi =\Phi_x+\Phi_{\Psi}\Psi_x+\Phi_{n}n_x+\Phi_{\rho}\rho_x+\Phi_qq_x$
is the total derivative and $\Phi_x =\partial_x\Phi$, etc. 

Suppose $\Delta_\nu(x,u^{(n)})=0,\;\nu =1,...,l,$ is a system of
differential equations, where 
$u^{(n)}=\left(\Psi, n, \rho, q, \Psi_x,\Psi_t,...,q_{tt}\right)$.
If ${\cal G}$ is a local
group of transformations acting on ${\cal M}$ and 
$pr^{(n)}\vec {{\cal V}}\left[\Delta_\nu(x,u^{(n)})\right]=0,\;\nu =1,...,l,$ 
whenever $\Delta(x,u^{(n)})=0$, 
for every infinitesimal generator $\vec {\cal V}$ of 
${\cal G}$, then ${\cal G}$ is a symmetry group of the system.

The standard 
procedure\footnote{A good description can be found in \cite{olver}} 
is based on finding the infinitesimal coefficient functions 
$\xi ,\tau ,\Phi, \Sigma ,\Gamma $ and $\Omega $.
Substituting the general formulae for $\Phi^x,\Sigma^x,etc.$ and equating
the coefficients of various monomials in the first and second order partial
derivatives of $\Psi, n, \rho$ and $q$, we find the defining equations. 
We wish to determine all possible coefficient functions 
$\xi ,\tau ,\Phi ,\Sigma ,\Gamma$
and $\Omega$ by solving the defining equations system so that the
corresponding one-parameter group $\exp (\varepsilon \vec {{\cal V}})$ 
will be the symmetry group of the equations \eref{eqs}. 
These solutions are:
\begin{equation}
\eqalign{
\fl{\bf Eckart:}\> 
\tau = c_1 + t c_4,\> \xi = c_2 + x c_4,\> \Phi = 0,
\> \Sigma = - n c_4,\> \Gamma = \rho c_3,\> \Omega = q c_3 \\
\fl {\bf Israel-Stewart:}\>
\tau = c_1 + t c_4,\> \xi = c_2 + x c_4,\> \Phi = 0, \> \Sigma = 0,
\> \Gamma = \rho c_3,\> \Omega = q c_3  }
\end{equation}
where $c_i$ are constants. The basis of the corresponding Lie algebra are:
\begin{equation}
\eqalign{
\fl {\bf Eckart:}
\>V_1=\partial_t,\> V_2=\partial_x,\> V_3=\rho\partial_{\rho}+q\partial_q,\> 
V_4=t\partial_t+x\partial_x-n\partial_n \\
\fl {\bf Israel-Stewart:}\>
V_1=\partial_t,\> V_2=\partial_x,\> V_3=\rho\partial_{\rho}+q\partial_q,\> }
\end{equation}

\section{Solvable group and invariants}\label{inv}
Because we have the Lie algebra of the equations \eref{eqs} we want to know if
the general solution of the system of differential equations can be found by
quadratures. This thing is possible if the Lie group is solvable. 
The requirement for solvability is equivalent to the existence of a basis 
$\left\{ {V_1,\ldots ,V_r}\right\} $ of Lie algebra ${\it g}$ such that 
\begin{equation}
[V_i,V_j]=\sum\limits_{k=1}^{j-1}c_{ij}^kV_k 
\end{equation}
whenever $i<j$.
Looking at the commutator table of the Lie algebra we will see that the
requirement of solvability is fulfilled for both Eckart and Israel-Stewart 
theories.
\vskip -4mm
\begin{table}[h]
\begin{center}
\caption{Commutator table for the Eckart and Israel-Stewart algebra}
\begin{tabular}{crrrr} \br 
{\bf [\, , \,]} & $V_1$ & $V_2$ & $V_3$ & $V_4$ \\  \mr
$V_1$ & 0 & 0 & $V_1$ & 0 \\
$V_2$ & 0 & 0 & $V_2$ & 0 \\
$V_3$ & -$V_1$ & -$V_2$ & 0 & 0 \\
$V_4$ & 0 & 0 & 0 & 0 \\ \br 
\end{tabular}
\hspace*{1cm}
\begin{tabular}{crrr} \br
{\bf [\, , \,]} & $V_1$ & $V_2$ & $V_3$ \\ \mr
$V_1$ & 0 & 0 & $V_1$  \\
$V_2$ & 0 & 0 & $V_2$  \\
$V_3$ & -$V_1$ & -$V_2$ & 0  \\ 
 & & & \\ \br
\end{tabular}
\end{center}
\end{table}
\vskip -8mm
We use the method of characteristics to compute the invariants of the Lie
algebra hopping that the reduced system, which can be obtained using the
invariants of the group, will help us to solve the system of equations 
\eref{eqs}. An n-th order differential invariant of a group G is a smooth
function depending on the independent and dependent variables and their
derivatives, invariant on the action of the corresponding n-th prolongation
of G \cite{olver}.
Suppose that we have the following generator: 
\begin{equation}
V_i=\tau_i\partial_t+\xi_i\partial_x+\Phi_i\partial_{\Psi}
+\Sigma \partial_n+\Gamma_i\partial_{\rho}+\Omega_i\partial_q 
\end{equation}
A local invariant $\zeta $ of $V_i$ is a solution of the linear, homogeneous
first order partial differential equation: 
\begin{equation}\label{invsol}
V_i(\zeta )=\tau_i\partial_t\zeta+\xi_i\partial_x\zeta 
+\Phi_i\partial_{\Psi}\zeta +\Sigma_i\partial_n\zeta 
+\Gamma_i\partial_{\rho}\zeta+\Omega_i\partial_q\zeta =0 
\end{equation}
The classical theory of such equations shows that the general solution of
equation \eref{invsol} can be found by integrating the corresponding
characteristic system of differential equations, which is 
\begin{equation}
\frac{dt}{\tau_i}=\frac{dx}{\xi _i}=\frac{d\Psi}{\Phi_i}
=\frac{dn}{\Sigma_i}=\frac{d\rho}{\Gamma_i}=\frac{dq}{\Omega_i} 
\end{equation}
Doing this integration we get, in this case, five invariants; we now
re-express the next generator of Lie algebra in terms of these five
invariants and then we perform another integration. We continue this
calculation until we re-express and integrate the last generator; at this
point we obtain a set of invariants that represent the system of independent
invariants of this group. The system of invariants can be used to reduce the
order of the original equations - constructing the reduced order system of
equations. Doing this one can hope to find simple equations that can be
integrated (for example \cite{olver}).

\section{Classification of group-invariant solutions}\label{clas}

A solution of the system of partial differential equations is said to be 
${\cal G}$-invariant if it is unchanged by all the group transformations in 
${\cal G}$. In general, to each s-parameter subgroup ${\cal H}$ of the full
symmetry group ${\cal G}$ of a system of differential equations, there will
correspond a family of group-invariant solutions. Since there are almost
always an infinite number of such subgroups, it is not usually feasible to
list all possible group-invariant solutions of the system. We need an
effective systematic means of classifying these solutions, leading to an
optimal system of group-invariant solutions from which every other solution
can be derived. Since the elements ${\it g}$ $\in $ ${\cal G}$ not in the
subgroup ${\cal H}$ will transform an ${\cal H}$-invariant solution to some
other group-invariant solution, only those solutions not so related need 
to be listed in our optimal system.
An optimal system of s-parameter subgroups is a list of conjugancy
inequivalent s-parameter subgroups with the property that any other subgroup
is conjugate to precisely one subgroup in the list (conjugancy map: 
$h\rightarrow ghg^-1$).
Let ${\cal G}$ be a Lie group with Lie algebra ${\it g}$; for each 
$v\in {\it g}$, the adjoint vector $ad\ v$ at $w\in $ ${\it g}$ is 
$ ad\ v|_w=[w,v]=-[v,w]. $
If $v\in {\it g}$ generates the one-parameter subgroup 
${\cal H}=\left[ \exp(\varepsilon v) : \varepsilon \in \Re \right]$, 
then $Ad\ g(v)$ will generate the conjugate one-parameter subgroup 
$g{\cal H}g^{-1}$.
Now we can construct the adjoint representation $Ad\ {\cal G}$ of the Lie
group by summing the Lie series 
\begin{equation}
Ad(\exp (\varepsilon v))w=\sum\limits_{n=0}^\infty \frac{\varepsilon ^n}{n!}%
(ad\ v)^n(w)=w-\varepsilon [v,w]+\frac{\varepsilon ^2}2[v,[v,w]]-... 
\end{equation}
obtaining the adjoint table. 
\vskip -5mm
\begin{table}[h]
\begin{center}
\caption{Adjoint table for Eckart and Israel-Stewart algebra.}
\begin{tabular}{crrcr} \br
{\bf Ad}  & {\bf $V_1$} & {\bf $V_2$} & {\bf $V_3$} & {\bf $V_4$}  \\ \mr
{\bf $V_1$} & $V_1$ & $V_2$ & $V_3-\varepsilon V_1$ & $V_4$ \\
{\bf $V_2$} & $V_1$ & $V_2$ & $V_3-\varepsilon V_2$ & $V_4$ \\
{\bf $V_3$} & $e^{\varepsilon}V_1$ & $e^{\varepsilon}V_2$ & $V_3$ & $V_4$ \\
{\bf $V_4$} & $V_1$ & $V_2$ & $V_3$ & $V_4$ \\ \br
\end{tabular}
\hspace*{1cm}
\begin{tabular}{crrc} \br
{\bf Ad}  & {\bf $V_1$} & {\bf $V_2$} & {\bf $V_3$}  \\ \mr
{\bf $V_1$} & $V_1$ & $V_2$ & $V_3-\varepsilon V_1$  \\
{\bf $V_2$} & $V_1$ & $V_2$ & $V_3-\varepsilon V_2$  \\
{\bf $V_3$} & $e^{\varepsilon}V_1$ & $e^{\varepsilon}V_2$ \\ 
 & & & \\ \br
\end{tabular}
\end{center}
\end{table}
\vskip -7mm
Finding the optimal system is a well known standard method \cite{olver};
we have found an optimal system of one-dimensional ($a\in\Re$) sub-algebras 
to be that spanned by:
\begin{equation}\label{otr}
\begin{array}{rll}
  &{\bf Eckart} & {\bf Israel-Stewart} \\
1)&\partial_x   & \partial_x \\
2)&\partial_t   & \partial_t \\
3)&x\partial_x+t\partial_t-n\partial_n &x\partial_x+t\partial_t-n\partial_n \\
4)&\partial_t+a\partial_x & \partial_t+a\partial_x \\
5)&x\partial_x+t\partial_t-n\partial_n+a\rho\partial_{\rho}+aq\partial_q & \\
6)&\rho\partial_{\rho}+q\partial_q+a\partial_x + \partial_t & \\
7)&\rho\partial_{\rho}+q\partial_q+\partial_x & \\
8)&\rho\partial_{\rho}+q\partial_q+\partial_t & \\
9)&\rho\partial_{\rho}+q\partial_q & 
\end{array}
\end{equation}

\section{Group-invariant solutions study}\label{sol}
Each symmetry of the optimal system of transformations \eref{otr} 
will be treated separately.

\begin{figure}[h]
\begin{center}
\epsfig{file=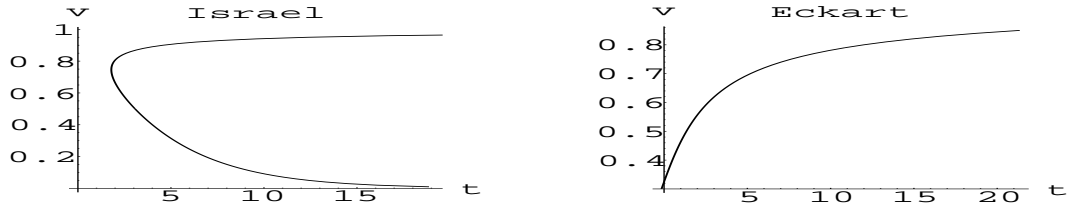,width=15cm,height=3cm}
\caption{Israel-Stewart and Eckart velocity solution, 
where t is $\frac{4kN_0}{\chi}t$.}
\label{f1}
\end{center}
\end{figure}

1) The fluid is spatially homogeneous and because the reduced system is very
simple the fluid equations can be immediately integrated analytically.
The \Fref{f1} shows that, for any initial state, the Eckart theory is 
unstable: as $t\rightarrow \infty$ then 
$v\rightarrow 1 \Rightarrow T\rightarrow \infty,\>n\rightarrow 0,\> 
\rho=3p\rightarrow 3E_0$ and $q\rightarrow -2E_0$.
The same thing happens, for the Israel-Stewart theory, only if the initial 
velocity is higher than the critical value $0.76$ but it can be stable for
initial values $v<0.76$, where $v=\tanh\Psi$. 
This spatially homogeneous case was also studied by Hiscock and Lindblom 
\cite{h4} and they found the critical value around $v_c=0.51$.

\begin{figure}[h]
\begin{center}
\epsfig{file=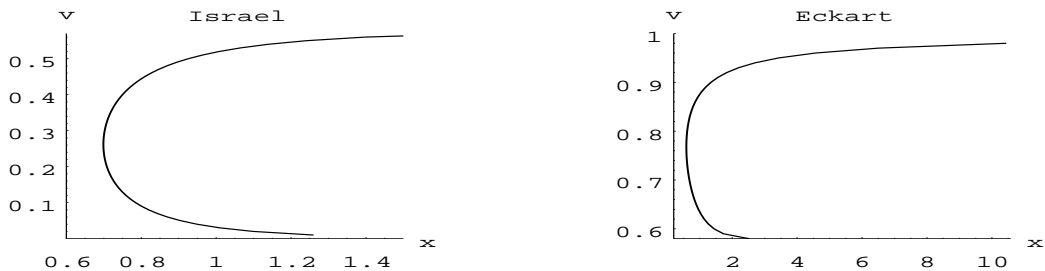,width=15cm,height=4cm}
\caption{Israel-Stewart and Eckart velocity solution, 
where x is $\frac{4kN_0}{\chi}x$.}
\label{f2}
\end{center}
\end{figure}

2) Also in this stationary case, the reduced system is very easy to integrate
and the space dependence of the velocity can be observed in the \Fref{f2}.
For the Eckart fluid, if the initial velocity is higher then 0.77 then as 
$x\rightarrow\infty$ we have $v\rightarrow 1 \Rightarrow T\rightarrow \infty,
\>n\rightarrow 0,\> \rho=3p\rightarrow 3E_0,\>q\rightarrow -2E_0$.
If we correlate this stationary behaviour with the previous result, 
we can comment that the stationary solution will increase in 
time for any initial Eckart velocity conditions; the Stewart-Israel fluid
has the chance to decrease when $t\rightarrow\infty,\>x\rightarrow\infty$
for a initial velocity smaller that $0.76$.

\begin{figure}[h]
\begin{center}
\epsfig{file=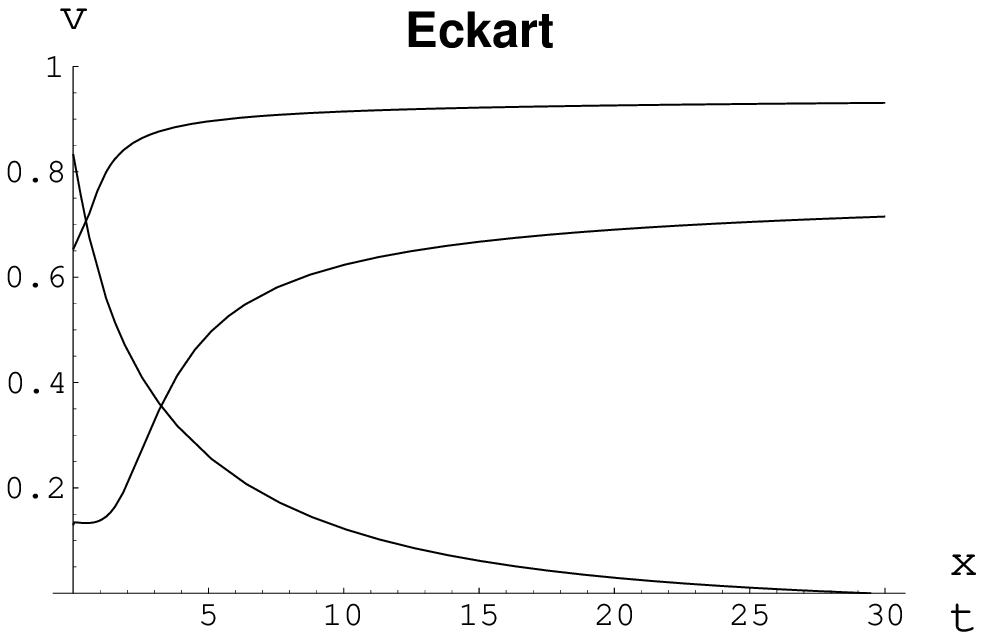,height=4cm}
\hspace*{1cm}
\epsfig{file=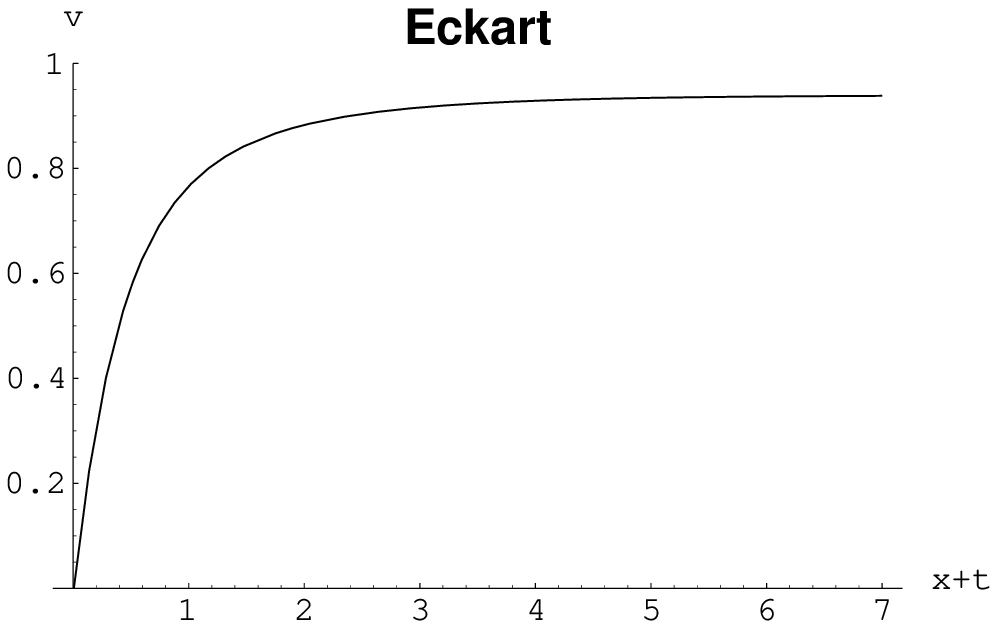,height=4cm}
\caption{Eckart velocity solution for the $3^{rd}$ and $4^{th}$ 
optimal transformations.}
\label{f34}
\end{center}
\end{figure}

3) The invariants of the characteristic equation $dx/x=dt/t=-dn/n$
are $\alpha=n\cdot t,\> y=x/t$ and for the Eckart fluid the reduced system of 
equations take the following form:
\begin{equation}
\eqalign{
\fl \alpha^{-1}\alpha_y=
-\frac{d}{dy}\left[\ln\left(\sinh\Psi+y\cosh\Psi\right)\right] 
\Rightarrow\alpha= N_0 \left(\sinh\Psi+y\cosh\Psi\right)^{-1} \\
\fl q=\frac{\rho_y}{6\Psi_y}
\frac{(1+y^2)\cosh(2\Psi)+2(-1+y^2+y\sinh(2\Psi)}{1-y^2}  \\
\fl 3q_y+4\rho\Psi_y+\rho_y
\frac{2y\cosh(2\Psi)+(1+y^2)\sinh(2\Psi)}{1-y^2}=0 \\
\fl \rho_y-\frac{3kN_0}{\chi}\frac{q}{\sinh\Psi+y\cosh\Psi}
+2\rho\frac{1+\left[(1+y^2)\cosh(2\Psi)+2y\sinh(2\Psi)\right]\Psi_y}{(1
+y^2)\sinh(2\Psi)+2y\cosh(2\Psi)} =0 }
\end{equation}
where $\Psi_y=\frac{d\Psi}{dy}$,etc.. 
The system of equations can be decoupled and the behaviour of the velocity 
numerical solution can be observed in the first picture of the \Fref{f34}.

It is important to mention that the first picture of the \Fref{f34}
shows that we have both exponentially increasing and decreasing solutions,
which means that we found a good Eckart velocity solution.
Also important is that if the initial velocity is smaller than $0.75$ we have
only decreasing solutions.
Further numerical investigations of the equation are needed in 
order to obtain information about the stability, initial condition 
dependence, etc..

For the Israel-Stewart fluid the reduced system is much
more complicated and we didn't succeed to solve the equations.

4) For this optimal transformation the invariant is $y=t-ax$; if we choose 
$a=-1$ the reduced system can be easily decoupled and for the Eckart fluid 
the solutions are: 
\begin{equation}
\eqalign{
v=\tanh\left[\ln\left[\sqrt{\chi C_1/(k N_0)}
\tanh\left(\sqrt{\chi C_1/(k N_0)}(y-C_2)\right)\right]\right] \\
n=N_0\sqrt{(1+v)/(1-v)},\>\> \rho_y=0,\>\>
q=\frac{2\chi}{3kN_0}\rho\frac{v_y}{(1+v)^2} }
\end{equation}
The velocity solution is ploted in the second picture of the \Fref{f34}. 
This $t+x$ dependence is typical for a wave type solution but 
as $y=t+x\rightarrow\infty$ we arrive again to wrong physical situation:
$v\rightarrow 1\Rightarrow n\rightarrow\infty,\>T\rightarrow 0$. 
The choice $a=-1$ does not affect the physics, it can change the $x$ axes 
direction or it can scale the space coordinate. 

The Israel-Stewart reduced system is also integrable but we found 
complex velocity solution.

5) The invariants are 
$\beta=n\cdot x,\> y=x/t,\>w=\rho\cdot t^a,\>\theta=q/\rho$. 
After some calculations, the reduced system of equations becomes:
\begin{equation}\label{o5}
\eqalign{
\fl \beta= N_0 y/(\sinh\Psi+y\cosh\Psi) \\
\fl w^{-1}w_y=
-\{ \left(\ln\left[(\sinh\Psi+y\cosh\Psi)(\cosh\Psi+y\sinh\Psi)\right]
\right)_y +\frac{(a-2)\sinh\Psi }{\cosh\Psi+y\sinh\Psi} \\
\fl
+ \frac{a((1+y)\sinh(2\Psi)+2y)}{(\sinh\Psi+y\cosh\Psi)(\cosh\Psi+y\sinh\Psi)
(6(y^2-1)\Psi_y-3a)} \} \\
\fl \left[1+\frac{3kN_0}{\chi}
\frac{2y\cosh(2\Psi)+(1+y^2)\cosh(2\Psi)+2(y^2-1)}{(\sinh\Psi+y\cosh\Psi)
(\cosh\Psi+y\sinh\Psi)(6(y^2-1)\Psi_y -3a)} \right]^{-1} \\
\fl \theta=
\frac{\left[w^{-1}w_y\left(2y\cosh 2\Psi+(1+y^2)\cosh 2\Psi+2(y^2-1)\right)
+a\left( (1+y)\sinh 2\Psi+2y\right) \right]}
{\left(6(y^2-1)\Psi_y-3a\right)} \\
\fl \frac{w_y}{w}\left(y\sinh 2\Psi+\cosh 2\Psi-2-3y\theta\right)-3y\theta_y
+a\left(\sinh 2\Psi-3\theta\right)-2\left(2y+3\theta\right)\Psi_y=0 }
\end{equation} 
The system of equations can be decoupled and solved numerically. 
The behaviour of the numerical velocity solution is the same as that was 
founded for the third optimal transformation; ofcourse, more precise 
numerical investigations are also needed.   

6) In this case the invariants are 
$y=x\pm at,\>\sigma=\rho\exp{(-t)},\>\alpha=q/\rho$ and the reduced system is:
\begin{equation}
\fl
\begin{array}{rl}
n_y \left(\sinh\Psi-a\cosh\Psi\right)+n\left(\cosh\Psi-a\sinh\Psi\right)\Psi_y
&=0\\
\sigma^{-1}\sigma_y\left(a\cosh\Psi-\sinh\Psi+a\alpha\sinh\Psi
-\alpha\cosh\Psi\right)+\alpha_y\left(a\sinh\Psi-\cosh\Psi\right) \\
+\cosh\Psi+\alpha\sinh\Psi+\left[4/3\left(a\sinh\Psi-\cosh\Psi\right)
+2\alpha\left(a\cosh\Psi-\sinh\Psi\right)\right]\Psi_y &=0\\
\sigma^{-1}\sigma_y\left(1/3\cosh\Psi-1/3a\sinh\Psi+\alpha\sinh\Psi
-a\alpha\cosh\Psi\right) \\
+\alpha_y \left(\sinh\Psi-a\cosh\Psi\right)-1/3\sinh\Psi-\alpha\cosh\Psi \\
+\left[4/3\left(\sinh\Psi-a\cosh\Psi\right)
+2\alpha\left(\cosh\Psi-a\sinh\Psi\right)\right]\Psi_y &=0\\
\cosh\Psi\left(\rho^{-1}\rho\sigma^{-1}\sigma-n^{-1}n_y \right)
-\sinh\Psi\left[\rho^{-1}\rho\left(a\sigma^{-1}\sigma+1\right)
-n^{-1}n_y \right] \\
+\left(\sinh\Psi-a\cosh\Psi\right)\Psi_y +3n\alpha k/\kappa &=0
\end{array}
\end{equation}
We can decouple the equations and in the end we obtain the following equation:
\begin{equation}\label{v6}
\Psi_{yy}+k_1\Psi^2_y+k_2f\left(\Psi\right)\Psi_y+k_3g\left(\Psi\right)=0
\end{equation}
where $k_{1,2,3}$ are constants and 
$f\left(\Psi\right),\>g\left(\Psi\right)$ are known functions of $\Psi$.
The equation \eref{v6} is integrable \cite{kamke}; because the integration 
algorithm is very long we will present only the final form of the equation
that can be integrated:
\begin{equation}\label{s6}
d\Psi(y)/dy=-t\xi'(t)\exp{(C_2\Psi)}
\end{equation} 
where $\xi=C_4\exp\left[(2-C_2)\Psi\right]$ and $C_i$ are constants; 
$t$, which is a function of $\Psi$ via $\xi$, has to be found from the 
equation $t^2\xi''(t)+C_4\xi(t)=0$. 
Because the analytical solution formula is very long and complicated 
we don't write it here, we prefer to mention that it has the same 
exponentially increasing trouble founded before.

For the 7),8) and 9) transformation we can't reduce the fluid system of 
equations.

\section{Summary and outlook \label{sum}}
The Lie symmetry group method was applied to two different relativistic 
heat-conducting fluid theories, the Eckart theory and the Israel-Stewart
theory, and a comparison between them has been made. We may summarise the
results as follows:
\begin{itemize}
\item applying the Lie symmetry group method to both fluids, we found that 
the Lie algebra of the Eckart theory has an additional transformation, 
namely the space, time and particle density dilatation $\left( V_4\right)$.
\item a systematic classification of the solutions is done by constructing
the optimal system of transformations. 
\item solving all the Eckart's reduced systems of equations, we found that
during the evolution of the fluid, there are, 
beside the exponentially increasing velocities, also solutions with 
good physical behaviour; we mention that in the 1 
dimensional approximation, for the Eckart theory, there are only 
exponentially increasing solutions.
\item due to a different (more complicate) phenomenological law, 
the optimal transformation system of the Israel-Stewart theory 
is trivial and as a consequence the reduced systems of equations are
not very much simplified but the transformation invariants; 
anyway, we found good physical behaviour for the spatially homogeneous
and stationary solutions.
\end{itemize}

Due to the fact that each fluid has a different phenomenological law,
the corresponding symmetry groups is different. We considered here only
two values for the parameter $\lambda$ that leads us to some well
known fluids, but performing the same kind of analysis for other $\lambda$
values, we could shed some light on the domain in which the relativistic 
heat-conducting fluid theory is well behaved.
Taking into account the bulk stress and the shear stress, the 
phenomenological law becomes more general but then the fluid equations will
be more difficult to solve and in spite of the tremendous work needed,
it is possible to obtain a very simple symmetry group that does not help
very much on the reduced system integration. 
It is important to point out that we are looking to find the simplest 
well behaved fluid theory in order to be solved much easier.
Anyway, it is an open subject that needs some attention. 

Because the 1+1 dimensional approximation is appropriate to describe the
collision of two highly Lorentz contracted heavy ions, further analysis
of the relativistic fluid, with or without viscosities, with different
$\lambda$ values, etc., are very important because, for some years, any
theoretical predictions can be compared with the experimental data 
that were taken at AGS or CERN. 
The concept of a quark-gluon plasma (QGP) predicted for heavy ion
collisions \cite{muller} has been the driving force for the experimental
and theoretical studies of hadronic matter at high energy densities. 
A lot of theoretical work has been done 
in order to use the hydrodynamical models in heavy ion physics and
several models based on relativistic hydrodynamics exist, 
the reader is reffered to Refs. \cite{csernai,str1,bjorken}. 

\ack
I wish to thank prof. M. Vi\c sinescu for continue encouraging this work.

\end{document}